\documentclass[aps,prl,twocolumn,showpacs]{revtex4}

\usepackage{graphicx,amssymb,amsfonts,amsmath,ifthen,bm}

\begin{document}

\title{Pairing and density-wave phases in Boson-Fermion mixtures at fixed
filling}
\author{F. D. Klironomos}
    \email{fkliron@physics.ucr.edu}
    \affiliation{Physics Department, University of California, Riverside,
CA 92521}
\author{S.-W. Tsai}
    \email{swtsai@physics.ucr.edu}
    \affiliation{Physics Department, University of California, Riverside, 
CA 92521}

\date{\today}


\newcommand{\ksif}[1]{\xi^f_{\bm #1}}
\newcommand{\ksib}[1]{\xi^b_{\bm #1}}

\begin{abstract}

We study a mixture of fermionic and bosonic cold atoms on a two-dimensional 
optical lattice, where the fermions are prepared in two hyperfine (isospin) 
states and the bosons have Bose-Einstein condensed (BEC). The coupling between 
the fermionic atoms and the bosonic fluctuations of the BEC has similarities 
with the electron-phonon coupling in crystals. We study the phase diagram for 
this system at fixed fermion density of one per site (half-filling). We find 
that tuning of the lattice parameters and interaction strengths (for 
fermion-fermion, fermion-boson and boson-boson interactions) drives the system 
to undergo antiferromagnetic ordering, s-wave and d-wave pairing 
superconductivity or a charge density wave phase. We use functional 
renormalization group analysis where retardation effects are fully taken into 
account by keeping the frequency dependence of the interaction vertices and 
self-energies. We calculate response functions and also provide estimates of 
the energy gap associated with the dominant order, and how it depends on 
different parameters of the problem.

\end{abstract}

\maketitle

Recent advances in manipulating ultracold atoms in optical lattices have 
produced remarkable results spanning a variety of modern condensed matter 
physics phenomena such as the realization of the Mott-insulator 
transition\cite{Greiner2002}, fermionic superfluidity\cite{Greiner2003}, 
the realization of the Tonks-Girardeau gas\cite{Paredes2004}, 
and the Berezinskii--Kosterlitz-–Thouless transition\cite{Hadzibabic2006},
just to name a few indicative ones.
The experimental control over a wide range of parameters associated with these
type of systems renders them a favorite physics playground both for 
experimentalists and theorists alike. In particular, a boson-fermion mixture
(BFM) system can provide a fascinating testing ground for a large section of 
theoretical physics\cite{Lewenstein2006} furthering our understanding of 
condensed matter phenomena.

Previous work has explored part of the vast and rich phase space associated with
this two-species system by employing mean-field or variational type of 
approaches and allowing for fluctuating occupation numbers for both fermions 
and bosons\cite{Buchler2004,Lewenstein2004,Illuminati2004,Wang2005,Sengupta2006}. 
As expected, these approaches provide a good general picture of the rich 
phase-space but are unable to capture the delicate interplay and competition 
of correlation effects in a quantitative manner which can be a useful reference
for experimentalists.
A more recent work, based on our methodology of analysis\cite{Mathey2006}, was 
the first approach to study this type of system beyond mean-field limitations 
but (as in all previous studies) was applied only for the regime $v_F\ll v_s$,
where retardation effects between the fermionic dynamics dictated by the Fermi 
velocity $v_F$ and the bosonic dynamics dictated by the sound velocity $v_s$ 
are not important.
It is very important to investigate the physics beyond this regime since
$v_F \geq v_s$ is experimentally accessible and in fact, retardation can play a
crucial role in the interplay of orders. The above considerations have prompted
us to focus our attention on this type of system and complement on the physical 
understanding of it by applying our theoretical apparatus originally developed
for the electron-phonon type of system\cite{SWTsai2005}.

The framework under which we work is based on functional renormalization group
(fRG) analysis for two-dimensional fermions\cite{Shankar1994} in the presence 
of phonons when full retardation is taken into 
consideration\cite{Klironomos2006}.
This approach has the important advantage of including all competing fermionic 
processes under a given (one-loop) accuracy approximation in an unbiased way, 
so that all possible orders are studied on an equal basis, contrary to mean 
field theory for example which always presupposes a manifested order.
The only necessary ingredient that is required for this study to work is the
bosonic system to be in the Bose-Einstein condensation phase (BEC) which 
leads to an electron-phonon type of fermion-boson interaction (linear to first
order in the bosonic field)\cite{Mathey2006}. 
This allows us to integrate the bosons out and build an effective fermionic
theory, and consequently apply the fRG analysis to investigate all possible
orders of interest the fermions can undergo, while taking retardation effects
fully into account.
What is interesting to realize is that our study can be generalized for any
type of a two-component itinerant fermionic system on a square lattice in the
presence of a gapless collective mode that renormalizes attractively the
fermionic interaction. All of our results will remain valid within that
context.

The standard model used to study itinerant fermions and bosons on a lattice is 
the Hubbard model. For a typical BFM system consisting of neutral atoms
(${}^{40}$K for fermions and ${}^{87}$Rb for bosons) that interact through
short-range van der Waals forces a good approximation consists 
only of hardcore (onsite) intra-species (fermion-fermion $U_{ff}$, 
boson-boson $U_{bb}$) repulsive interactions and inter-species 
(fermion-boson $U_{fb}(\bm q)$) repulsive or attractive interactions.
Additionally, the presence of a two-component isospin quantum number among 
fermions relaxes Pauli's exclusion principle limitations for s-wave type of 
scattering events (lowest in energy) and opens up a large and rich
phase space previously inaccessible for ``spinless" fermionic atoms. This type
of two-component fermionic system in the presence of a collective mode can 
effectively map and simulate a large variety of correlated electronic systems 
in condensed matter physics\cite{Lewenstein2006}.
The role of isospin in this case is played by different total angular momentum
projection states for the ${}^{40}$K atoms, which can additionally be exploited
for Feshbach resonance type of tuning of the inter-species scattering strength 
$U_{fb}$\cite{Ferlaino2006}.

The Hamiltonian for this system of itinerant fermions and bosons on a 
two-dimensional square lattice has the form
\begin{eqnarray}
&H&\!\!=\sum_{\bm k,\sigma}\ksif{k}f^\dagger_{\bm k,\sigma}f_{\bm k,\sigma} 
+ \sum_{\bm q}\ksib{q}B^\dagger_{\bm q}B_{\bm q} \notag\\
&+&\frac{1}{2V}\sum_{\bm k,\bm k',\bm q}U_{bb}B^\dagger_{\bm k+\bm q}B_{\bm k}
B^\dagger_{\bm k'-\bm q}B_{\bm k'}\notag\\
&+&\frac{1}{V}\sum_{\bm k,\bm k',\bm q,\sigma} U_{fb}(\bm q)
f^\dagger_{\bm k+\bm q,\sigma}f_{\bm k,\sigma}B^\dagger_{\bm k'-\bm q}
B_{\bm k'} \notag\\
&+&\frac{1}{V}\sum_{\bm k,\bm k',\bm q}U_{ff}f^\dagger_{\bm k+\bm q ,\uparrow}
f_{\bm k ,\uparrow}f^\dagger_{\bm k'-\bm q ,\downarrow}f_{\bm k' ,\downarrow},
\label{H_original}
\end{eqnarray}
where $f_{\bm k,\sigma},B_{\bm q}$ are the fermion, boson operators 
respectively, $\xi^{f,b}_{\bm k}=-2t_{f,b}(\cos{k_x}+\cos{k_y}) - \mu_{f,b}$
are the corresponding dispersion relations dictated by the square lattice 
symmetry, with $t_{f,b}$ being the overlap integrals and $\mu_{f,b}$ the 
chemical potentials for the fermions and bosons respectively.

For the purpose of this study we will focus our attention at half-filling for
the fermions ($\mu_f=0$) where the pairing-related phases (s,d-wave) 
superconductivity (sSC, dSC) can compete with the nesting-related phases 
like charge density wave (CDW) and antiferromagnetism (SDW), and a rich 
landscape of phase space becomes available for the fermionic system to 
explore\cite{Klironomos2006}. 
This amounts to fermion numbers in the range $10^3$--$10^4$ for typical optical
lattice parameters $a=400$nm on 2D optical harmonic traps of frequency 
$\omega=2\pi\times 30$Hz, which is well within experimental capabilities.
Additionally, at half-filling besides the presence of nesting on the Fermi
surface there are van Hove singularities as well that can enhance the interplay
and competition among all orders.
Our assumption of BEC for the bosons allows us to expand the bosonic operators 
in Eq.~(\ref{H_original}) according to 
$B_{\bm q}=\sqrt{N_0}\delta_{\bm q,0}+B_{\bm q}$, where the creation 
($B^\dagger_{\bm q}$) or destruction ($B_{\bm q}$) of quasiparticles follows
from fluctuations ($B_{\bm 0}$) of the macroscopic number $N_0$ of condensate
atoms. Consequently we can perform the usual Bogoliubov transformation 
and linearize the inter-species interaction introducing the gapless collective 
mode
\begin{eqnarray}
\omega_{\bm q}=\sqrt{\varepsilon_{\bm q}(\varepsilon_{\bm q}+2n_bU_{bb})},
\label{omega_q}
\end{eqnarray}
where $\varepsilon_{\bm q}=\ksib{q}{}_{,\mu_b=-4t_b}$ is the free-boson 
dispersion relation (chemical potential exactly at the non-interacting ground 
state), and $n_b=N_0/V$ is the BEC density which for all practical purposes
can be considered as the bosonic filling factor on the lattice.

This is the fixed point of our theory as far as the stability of the bosonic
system in the presence of fermions is concerned. It has been experimentally
shown\cite{Modugno2002,Roati2002,Ospelkaus2006} that the BEC can be stable in
the presence of fermionic atoms provided the two-species numbers do not exceed
$10^{5}$ atoms in the optical trap {\it in the absence of an optical lattice}. 
The presence of an optical lattice will stabilize the system against phase 
separation or collapse to even higher densities, since both processes rely on 
the inter-species interaction $U_{fb}$ dominating over the kinetic energy scales
associated with the weakly confined atoms, and three-body recombination 
processes becoming relevant as well. 
The lattice will introduce experimentally tunable kinetic energy scales 
associated with the hopping of atoms from lattice point to lattice point
($t_{f,b}$) competing with $U_{fb}$ and consequently stabilizing the system.

Dynamical fluctuations of the BFM around this fixed point can be modeled in the
Matsubara representation. After the bosonic fields associated with fluctuations 
of the BEC are integrated out\cite{Mathey2006,SWTsai2005} we arrive at the
fully retarded effective fermionic interaction given by
\begin{widetext}
\begin{eqnarray}
\widetilde{U}_{ff}(k_1,k_2,k_3)=U_{ff}-\frac{U^2_{fb}/U_{bb}}
{1+4\xi^2-2\xi^2\big(\cos{(\bm k_1-\bm k_3)_x}+\cos{(\bm k_1-\bm k_3)_y}\big)}
\frac{\omega^2_{\bm k_1-\bm k_3}}{(\omega_1-\omega_3)^2
+\omega^2_{\bm k_1-\bm k_3}},
\label{U_ff}
\end{eqnarray}
\end{widetext}
where $k_i\equiv(i\omega_{n_i},\bm k_i)$, $\omega_{n_i}$ are the fermionic
Matsubara frequencies, and $\xi=\sqrt{t_b/2n_bU_{bb}}$ is the healing length
of the BEC.
As we see, fluctuations of the BEC renormalize $U_{ff}$ with an attractive and
frequency-dependent part which is not affected by the attractive or repulsive 
nature of the inter-species interaction $U_{fb}$\cite{Illuminati2004}. 
The importance of retardation is determined by Eq.~(\ref{omega_q}) which in
the longwavelength limit defines the acoustic velocity 
$v_s=\sqrt{2t_bU_{bb}n_b}$. For fermionic velocities ($v_F\simeq 2t_f$) much 
smaller than $v_s$ one recovers the mean-field limit for 
$\widetilde{U}_{ff}$\cite{Illuminati2004}. 
On the other hand, retardation dominates for $v_s\leq v_F$ and the competition
of phases is enhanced. This regime can be achieved experimentally
since there are two tunable parameters ($t_b$, $U_{bb}$) that can be adjusted 
{\it in situ} for a given system configuration ($n_b$).
The healing length $\xi$ plays an important role in the competition of phases
since it defines the length-scale over which bosonic correlations are 
present\cite{Wang2005}.
\begin{figure}[t]
\includegraphics[angle=-90,totalheight=5.5cm,width=7.0cm,viewport=5 5 560
690,clip]{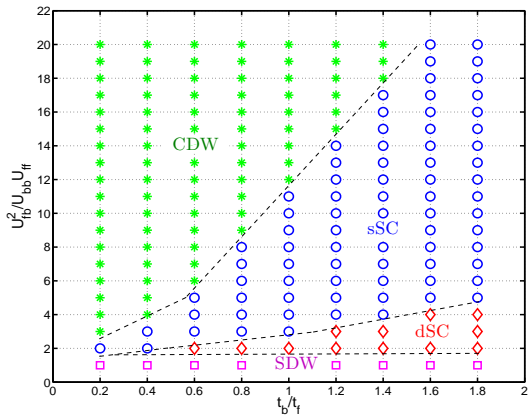}
\caption{(Color online) Phase diagram for $U_{ff}=0.4t_f$, $U_{bb}=0.8t_f$,
and $n_b=2.5$.
Blue circles indicate sSC, red rhombuses indicate dSC, magenta squares SDW, 
and green stars CDW type of ordering. Dashed lines are guides to the eye.}
\label{Fig:fully-retarded}
\end{figure}
If these correlations, which enter the attractive part of the fermionic 
interaction, are short-ranged $\xi\leq 1$ (in units of the lattice 
constant) this will favor CDW (at half-filling) competing with s-wave type
of pairing. 
If on the contrary $\xi>1$, the lattice symmetry becomes visible and exotic
pairing phases enter the competition\cite{Wang2005,Mathey2006} along with SDW
(at half-filling). This is reflected throughout our results as well as we 
discuss below.

Our fRG analysis for the fermionic system was applied at zero temperature for
the electron-phonon system at half-filling\cite{Klironomos2006}. 
It involves the self-consistent integration of fast-modes, determined by the
energy cutoff $\Lambda(\ell)=4t_fe^{-\ell}$ ($\ell$ is the RG-step in the
integration process) which renormalizes the fermionic interaction until a
divergence occurs as the limit $\ell\rightarrow\infty$ is approached.
At half-filling the presence of van Hove singularities on the Fermi surface 
allows us to focus our attention only on $\bm k$-space processes around
the singular points $\bm Q_1=(\pi,0)+\bm k$ and $\bm Q_2=(0,\pi)+\bm k$ with 
$|\bm k|\ll\pi$, called the two-patch approximation. As a result, a general
fermion-fermion coupling $U_{ff}(k_1,k_2,k_3)$ reduces in momentum-space to four
non-redundant couplings associated with the scattering processes of 
$g_1 \equiv U_{ff}(\bm Q_1,\bm Q_2,\bm Q_2)$,
$g_2 \equiv U_{ff}(\bm Q_1,\bm Q_1,\bm Q_1)$, 
$g_3 \equiv U_{ff}(\bm Q_1,\bm Q_1,\bm Q_2)$,
$g_4 \equiv U_{ff}(\bm Q_1,\bm Q_2,\bm Q_1)$, where we have suppressed 
the frequency dependence for clarity.
The different orders of interest are defined in terms of these 
couplings according to\cite{Schulz1987}
\begin{eqnarray}
u_{\textrm{(sd)SC}}&=&-2(g_2\pm g_3), \label{u:sdSC}\\
u_{\textrm{SDW}}&=&2(g_4+g_3), \label{u:SDW}\\
u_{\textrm{CDW}}&=&-2(2g_1+g_3-g_4), \label{u:CDW}
\end{eqnarray}
where the signs are chosen so that all orders diverge to positive values when
they strongly renormalize. In this two-patch approximation the effective
interaction of Eq.~(\ref{U_ff}) reduces to the pairs of
\begin{eqnarray}
\!\!\!\!\!\!&g_{1,3}&\!\!\!(\omega_1,\omega_2,\omega_3)\!=\!U_{ff}
-\frac{U_{fb}^2}{U_{bb}}\frac{8t_b^2/\xi^2}{(\omega_1\!-\!\omega_3)^2\!\!
+\!\!\frac{8t_b^2}{\xi^2}(1\!+\!8\xi^2)},\label{g13}\\
\!\!\!\!\!\!&g_{2,4}&\!\!\!(\omega_1,\omega_2,\omega_3)\!=\!U_{ff}
-\frac{U_{fb}^2}{U_{bb}}\delta_{\omega_1,\omega_3},\label{g24}
\end{eqnarray}
which are inherently {\it anisotropic} ($\delta_{\omega_1,\omega_3}$ is the 
Kronecker delta). This inherent anisotropy can enhance d-wave pairing for
example in the regime where $g_2<0, g_3>0$ with $|g_3|>|g_2|$ as 
Eq.~(\ref{u:sdSC}) shows.
Finally, in order to account for all frequency channels correctly one needs to 
evaluate the susceptibilities of the different orders and identify the most
divergent one\cite{Klironomos2006}.

We have performed our fRG analysis and have investigated various parameter 
ranges associated with the experimentally tunable set of 
$\{U_{ff},U_{fb},U_{bb},t_b,t_f,n_b\}$. As is reasonably expected, for 
$U_{fb}^2/U_{bb}U_{ff}<1$ the effective interaction of Eq.~(\ref{U_ff}) remains
repulsive and antiferromagnetic ordering dominates throughout the range of 
parameter space. Interesting physics arises when $U_{fb}^2/U_{bb}U_{ff}\geq 1$.
In Fig.~(\ref{Fig:fully-retarded}) we present the phase diagram our fRG analysis
produces for $U_{ff}=0.4t_f$, $U_{bb}=0.8t_f$ and $n_b=2.5$ and a range of 
$U_{fb}^2/U_{bb}U_{ff}>1$, $t_b/t_f$ values.
\begin{figure}[t]
\includegraphics[angle=-90,totalheight=5.5cm,width=7.0cm,viewport=5 5 560
690,clip]{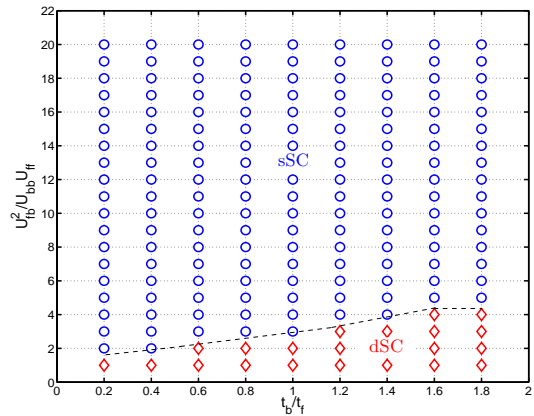}
\caption{(Color online) Phase diagram for $U_{ff}=0.4t_f$, $U_{bb}=0.8t_f$,
and $n_b=2.5$ in the limit $v_F\ll v_s$ where retardation is not important. 
In this case, the whole CDW phase disappears.
Symbols are same as in Fig.~(\ref{Fig:fully-retarded}) and dashed line is
guide to the eye.}
\label{Fig:no-retardation}
\end{figure}
We notice that even at half-filling a large region of the phase diagram is 
taken over by pairing related phases. This is the essential new result of this
work and the difference with previous studies, where mean-field\cite{Wang2005}
and $v_F<<v_s$ fRG\cite{Mathey2006} focused on different dopings for the
fermionic system. Both were unable to capture at half-filling the rich
interplay of orders and their strong dependence on $t_b$.
As $t_b$ is strengthened, the bosonic correlation length $\xi$ (which is 
independent of $U_{fb}$) becomes larger allowing for boson-mediated attraction
among the fermions to reach over many lattice sites increasing the Cooper
pair sizes and allowing fermions to avoid on-site pairing and the associated
energy cost of $U_{ff}$. For small $U_{fb}$ values this mechanism is
detrimental for CDW and only when the bosonic correlation length 
$\xi\leq 1/2$ can CDW win over s-wave pairing.
For larger values of $U_{fb}$ charge order can survive over sSC in a region
$1/2\leq\xi<1$. If the range of the attractive interaction exceeds the lattice 
constant value then pairing across different sites becomes energetically
favorable and CDW becomes subleading.

At this point we should mention that the $\xi\sim 1/2$ condition for CDW to
appear indicates that the BEC phase has weakened as a whole and long range
order has been lost.
In free space, the consequence of this is that our linear approximation of
coupling fermions to only the bosonic density fluctuations might not be
adequate, there will be lots of quasiparticle excitations.
Nevertheless, we believe that even if those processes are taken into account
the strong influence of the lattice and the presence of nesting on the Fermi
surface will sustain the CDW order to be experimentally observed.
\begin{figure}[t!]
\includegraphics[angle=-90,totalheight=5.5cm,width=7.0cm,viewport=5 5 560
700,clip]{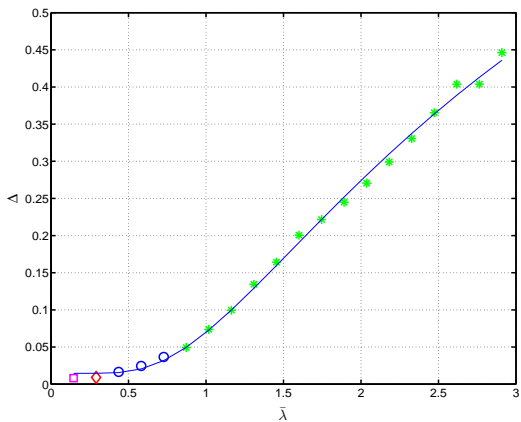}
\caption{(Color online) Evolution of the gap along $t_b/t_f=0.6$ of 
Fig.~(\ref{Fig:fully-retarded}) with identical symbol scheme for the different 
orders. The blue line fitting was according to 
$\Delta=0.015+3.326\exp(-(3.101+\bar{\lambda})/\bar{\lambda})$.}
\label{Fig:gap}
\end{figure}

As we mentioned in the introduction, all previous works on this type of system
have ignored retardation effects. This is an {\it a priori} assumption based on
comparing the free-space masses of ${}^{40}$K and ${}^{87}$Rb\cite{Wang2005}.
When the lattice is turned on, one should be comparing band-masses related to
$t_f$, $t_b$ which can be tuned to any range and consequently $v_F,v_s$ can be 
independently set experimentally. 
In Fig.~(\ref{Fig:no-retardation}) we show the same phase diagram as in 
Fig.~(\ref{Fig:fully-retarded}) without retardation taken into 
consideration, which effectively amounts to setting $\omega_1=\omega_3$ for all 
frequency channels in Eqs.~(\ref{g13}-\ref{g24}). We have recovered
the results at half-filling of the previous fRG study\cite{Mathey2006}. 
Antiferromagnetism is expected to take over at the $U_{fb}^2/U_{bb}U_{ff}<1$
region. As it can be clearly seen by contrasting the two phase diagrams,
retardation is very important at tuning the fermionic phases.

Our analysis has the additional advantage of qualitatively producing the 
energy gap associated with the dominant orders, $\Delta=4t_fe^{-\ell_c}$, just
by following the critical RG-step ($\ell_c$) at which they diverge.
We can define 
\begin{equation}
\bar{\lambda}=\frac{U_{fb}^2}{2U_{bb}}\frac{2+8\xi^2}{1+8\xi^2},
\label{lambda_bar}
\end{equation}
as the average anisotropic coupling strength of the fermion-boson interaction
(easily derived from Eqs.~(\ref{g13}-\ref{g24})) and investigate the 
functional form of $\Delta(\bar{\lambda})$.
In Fig.~(\ref{Fig:gap}) we show the evolution of the gap along the 
$t_b/t_f=0.6$ direction in Fig.~(\ref{Fig:fully-retarded}), where the 
fermionic system passes through all four phases. The functional form we use
for the fitting is $\Delta=d_1+d_2\exp(-(d_3+\bar{\lambda})/\bar{\lambda})$
with $d_1=0.015$, $d_2=3.326$, $d_3=3.101$.

In conclusion we have applied functional renormalization group method with
full retardation for a boson-fermion mixture system on an optical lattice when
the fermions are at half-filling and the bosons have Bose-Einstein condensed.
We find a rich phase diagram providing quantitative estimates for the 
experimental observation of the competing orders. Additionally, we have produced
a functional dependence of the energy gap of the dominant orders with the 
average boson-fermion coupling strength.

We would like to acknowledge fruitful discussions with Karyn Le Hur and Antonio
Castro Neto.

\end{document}